# Human-AI Interaction and User Satisfaction:
# Empirical Evidence from Online Reviews of AI Products

Stefan Pasch[1]    Sun-Young Ha[2]

**Abstract**

Human-AI Interaction (HAI) guidelines and design principles have become increasingly important in both industry and academia to guide the development of AI systems that align with user needs and expectations. However, large-scale empirical evidence on how HAI principles shape user satisfaction in practice remains limited. This study addresses that gap by analyzing over 100,000 user reviews of AI-related products from G2.com, a leading review platform for business software and services. Based on widely adopted industry guidelines, we identify seven core HAI dimensions and examine their coverage and sentiment within the reviews. We find that the sentiment on four HAI dimensions—adaptability, customization, error recovery, and privacy & security—is positively associated with overall user satisfaction. Moreover, we show that engagement with HAI dimensions varies by professional background: Users with technical job role are more likely to discuss system-focused aspects, such as reliability, while non-technical users emphasize interaction-focused features like customization and feedback. Interestingly, the relationship between HAI sentiment and overall satisfaction is not moderated by job role, suggesting that once an HAI dimension has been identified by users, its effect on satisfaction is consistent across job roles.

## 1. Introduction

With the rise of large language models (LLMs) and the growing integration of artificial intelligence (AI) systems into everyday work, designing these technologies with user interactions in mind has become increasingly important. In response, the concept of Human-AI Interaction (HAI) has emerged as a key concept in both industry and academia. HAI refers to the study and design of how humans engage with AI systems across a range of tasks, capabilities, and design contexts. Building on principles from human-computer interaction (HCI), artificial intelligence, and user-centered design, HAI research has emphasized transparency, usability, and alignment with human goals (Amershi et al., 2019; Yang et al., 2020; Eiband et al., 2018). Reflecting this, both researchers and practitioners have developed guidelines and best practices to promote more effective and trustworthy interactions with AI (Amershi et al., 2019; Rai, 2020; Shin, 2020).

While the conceptual relevance of HAI is widely recognized, empirical evidence on how these principles shape real-world user experiences remains limited. Much of the existing work is grounded in theoretical frameworks, literature reviews, expert interviews, or small-scale qualitative studies (e.g., Amershi et al., 2019; Shneiderman, 2020; Eiband et al., 2018). As a result, there is limited large-scale empirical evidence on how users engage with HAI principles in real-world settings, and to what extent these dimensions shape their satisfaction with AI systems and products.

[1] Division of Social Science & AI, Hankuk University of Foreign Studies: stefan.pasch@outlook.com
[2] sunnyoung.ha@gmail.com

This study addresses this gap by analyzing a large corpus of user-generated reviews of AI-related software products collected from G2.com, a leading online platform for business software. To examine how users discuss and evaluate Human-AI Interaction, we develop a novel measurement framework that links real-world user feedback to established HAI guidelines. Specifically, we extract key principles from three widely used industry frameworks – developed by Microsoft, Google, and Apple - and cluster them into seven overarching HAI dimensions: feedback, adaptability, customization, explainability, error recovery, reliability, and privacy & security.

We then apply modern natural language processing (NLP) techniques—including sentence embeddings and topic modelling using BERTopic—to analyze more than 100,000 user reviews. BERTopic is a state-of-the-art machine learning method that groups similar pieces of text into coherent topics, helping to identify the different themes that users discuss (Grootendorst, 2022). Out of these, we select the topics most closely related to each HAI dimension using semantic similarity measures. This allows us to (i) identify when HAI dimensions are discussed, (ii) determine whether they are mentioned positively or negatively (HAI sentiment), and (iii) assess how these evaluations on HAI relate to overall user satisfaction with the corresponding AI product, as reflected in review scores.

However, the link between HAI evaluations and overall satisfaction may not be uniform across users. AI systems are used by a diverse range of professionals—from technical experts such as data scientists and software engineers to non-technical users like marketers and content creators. These groups likely evaluate AI systems in different ways based on their needs and responsibilities.

This expectation aligns with insights from Human-Computer Interaction (HCI) research, which suggest that users' roles shape how they interact with technology (Grudin, 1991; Ko et al., 2006). In particular, technical professionals tend to prioritize system-level attributes such as transparency, robustness, and error handling (Ko et al., 2006; Carroll, 1997), while non-technical users focus more on usability, adaptability, and seamless integration into their workflows. Drawing on this literature, we distinguish between system-focused HAI dimensions (explainability, error recovery, reliability, and privacy & security) and interaction-focused dimensions (feedback, customization, and adaptability), and examine whether users' professional background influences which of these are emphasized.

To explore this, we use job title information provided in the reviews to group users into professional categories, allowing us to investigate how engagement with HAI dimensions varies across technical and non-technical job roles—and whether these roles also moderate the relationship between HAI sentiment and satisfaction.

Taken together, we address three key research questions:

1. Is user sentiment toward specific HAI dimensions associated with overall satisfaction?
2. Do different types of users (e.g., technical vs. non-technical professionals) emphasize different aspects of Human-AI Interaction?
3. Does the influence of HAI sentiment on satisfaction vary depending on the reviewer's professional background?

By combining conceptual insights from HAI and HCI research with large-scale user-generated data, this study makes the following key contributions:
- It provides empirical evidence linking sentiment on HAI dimensions to user satisfaction across a diverse range of AI tools.
- It introduces a scalable methodology for extracting and quantifying HAI-related themes in real-world user feedback.
- It explores how professional background shapes engagement with HAI topics—both in terms of which dimensions users emphasize and whether their roles moderate the relationship between HAI sentiment and user satisfaction.

Together, these findings offer both practical and theoretical implications for the development of more user-aligned, trustworthy AI systems.

## 2. Theoretical Background

**Human-AI Interaction and User Satisfaction**

The importance of Human-AI Interaction (HAI) principles is increasingly recognized across fields such as AI ethics, usability, and interaction design research. Industry guidelines, including those from major technology companies like Microsoft, emphasize that AI systems should be transparent, reliable, adaptable, and user-centric (Amershi et al., 2019). Similarly, academic literature highlights principles such as explainability, fairness, and personalization as crucial for fostering trust, acceptance, and usability of AI applications (Rai, 2020; Shin, 2020). These HAI principles and practices, if implemented successfully, are expected to enhance user satisfaction through several potential channels.

A key mechanism through which HAI can enhance user satisfaction is trust. HAI principles are believed to increase trustworthiness by ensuring that AI behaves predictably and aligns with user expectations (Shneiderman, 2020). Among these principles, explainability has been found to be particularly influential in fostering trust (Rai, 2020). Prior research suggests that trust, once established, in turn, significantly enhances user satisfaction with AI technologies (Cheong et al., 2023).

Similarly, HAI principles related to perceived ease-of-use and intuitiveness are designed to reduce the cognitive effort required from users, facilitating effortless interactions and enhancing usability (Amershi et al., 2019; Yang et al., 2020). Empirical research from the HCI literature robustly demonstrates that ease-of-use is a critical determinant of technology acceptance, adoption, and overall user satisfaction (Venkatesh et al., 2003; Davis, 1989). Thus, intuitive, user-centric AI designs that reduce user effort should directly contribute to increased satisfaction and positive user experiences.

Personalization is another example of an HAI principle that aims to align AI systems with individual user preferences, intending to enhance users' sense of agency and relevance during interactions (Amershi et al., 2019; Knijnenburg et al., 2012). Empirical evidence strongly supports that personalization significantly enhances users' perceived usefulness and satisfaction with AI technologies, due to increased alignment with individual user needs and preferences (Yeung et al., 2019; Knijnenburg et al., 2012).

Finally, HAI design practices that emphasize accessibility, fairness, and ethical alignment aim to foster user perceptions of trustworthiness, inclusivity, and broader acceptance. Existing research indeed confirms that ethically aligned and inclusive systems consistently achieve higher levels of user trust and satisfaction (Yang & Lee, 2024; Usmani et al., 2023; Pasch, 2025).

Given these considerations, we hypothesize:

*H1: Sentiment toward dimensions of Human-AI Interaction is positively associated with user satisfaction.*

**User Roles in Human-AI Interaction: System-Focused vs. Interaction-Focused Dimensions**

An increasing number of job roles interact with AI-driven technologies across diverse contexts, ranging from technical specialists and developers who design, evaluate, or maintain AI systems to non-technical professionals and everyday users who primarily rely on AI products to accomplish practical tasks (Yang et al., 2020).

The Human-Computer Interaction (HCI) literature has long suggested that users evaluate and prioritize interaction principles differently based on their roles, tasks, and expertise (Grudin, 1991; Ko et al., 2006). HCI research has emphasized that effective design must account for the distinct cognitive demands and interaction goals of different user groups (Dillon and Watson, 1996; Fischer, 2001). In particular, Grudin (1991) identifies two broad categories of users: (i) Developers (technical experts) – who engage deeply with system internals, prioritizing attributes like reliability, transparency, and error-recovery; and (ii) End-users (non-technical users) – who primarily evaluate technology based on ease of use, intuitiveness, and seamless integration into everyday workflows.

Related findings further suggest that developers and technical specialists emphasize system-focused attributes because their tasks inherently involve understanding, maintaining, and improving technological systems (Ko et al., 2006; Blackwell, 2002). Conversely, non-technical users prioritize usability, task completion, and intuitive interaction, as their focus lies in achieving practical goals rather than understanding the underlying technology (Carroll, 1997; Davis, 1989). This distinction aligns with perspectives in HCI research that differentiate between system-centered functionality and interaction-focused usability for end-users (Gasson, 2003; Díaz et al., 2008).

Given the broad range of user roles engaging with AI products (Yang et al., 2020), these role-based differences are particularly relevant for Human-AI Interaction. Recent research has drawn similar distinctions between system-oriented HAI dimensions (concerned with AI functionality, transparency, and reliability) and interaction-oriented HAI dimensions (focused on usability, adaptability, and interaction quality) (Eiband et al., 2018; Yang et al., 2020). For example, Microsoft's Human-AI Interaction (HAI) guidelines explicitly separate interaction-focused principles (e.g., optimizing user engagement "during interaction" and interactions "over time") from system-focused principles (e.g., ensuring and explaining AI's "initial" functionality and handling errors "when wrong"). Similarly, in this study, we adopt the distinction between system-focused and interaction-focused HAI dimensions to better understand how different aspects of AI impact user experience and satisfaction:

- **System-Focused HAI Dimensions** (e.g., interpretability, transparency, reliability, robustness) explicitly focus on supporting users' analytical understanding and evaluation of AI systems. These dimensions are essential for users who need to comprehend, predict, and potentially intervene in AI operations (Xiong et al., 2022; Lipton, 2016; Corso et al, 2023).
- **Interaction-Focused HAI Dimensions** prioritize enhancing user experience and interaction quality, including adaptability, personalization, and customization (Amershi et al., 2019; Sundar et al., 2024).

This distinction is particularly important in AI applications, where the complexity and black-box nature of AI models make system transparency essential for trust and accountability, while interaction quality ensures usability and effective human-AI collaboration (Yang et al., 2020). By integrating HCI research on role-based differences (Grudin, 1991; Fischer, 2001; Ko et al., 2006) with Human-AI Interaction literature (Amershi et al., 2019; Yang et al., 2020), we anticipate that technical and non-technical users will prioritize different aspects of AI.

- **Technical professionals**, such as engineers and data scientists, focus on system-oriented dimensions, as these factors directly affect their ability to understand, optimize, and troubleshoot AI systems (Xiong et al., 2022; Lipton, 2016).
- **Non-technical (end-)users** evaluate AI based on interaction-oriented dimensions prioritizing seamless integration into daily workflows and interaction quality over system-level transparency (Sundar et al., 2024; Davis, 1989).

These differences should be reflected in the frequency of discussions on HAI topics and should also affect the link between HAI practices and satisfaction. Technical users, whose work depends on understanding AI's internal mechanisms, are more likely to discuss system-level properties in their evaluations. Meanwhile, non-technical users, focused on practical outcomes, are more concerned with how well AI supports their tasks, making usability the dominant theme in their discussions.

Beyond discussion levels, the influence of HAI dimensions on satisfaction should also differ by user role. For technical users, system-oriented features like explainability and reliability directly impact their ability to trust and manage AI systems, making them critical drivers of satisfaction (Lipton, 2016). Non-technical users, however, derive satisfaction from smooth, intuitive interactions, meaning that usability and personalization are more influential than technical understansing (Venkatesh & Davis, 2000).

*H2a: Technical professionals discuss system-focused HAI dimensions more frequently and interaction-focused dimensions less frequently than non-technical users.*

*H2b: The effect of HAI dimensions on user satisfaction is moderated by professional background, such that system-focused dimensions have a stronger impact on satisfaction for technical users, while interaction-focused dimensions have a stronger impact on satisfaction for non-technical users.*

## 3. Methodology

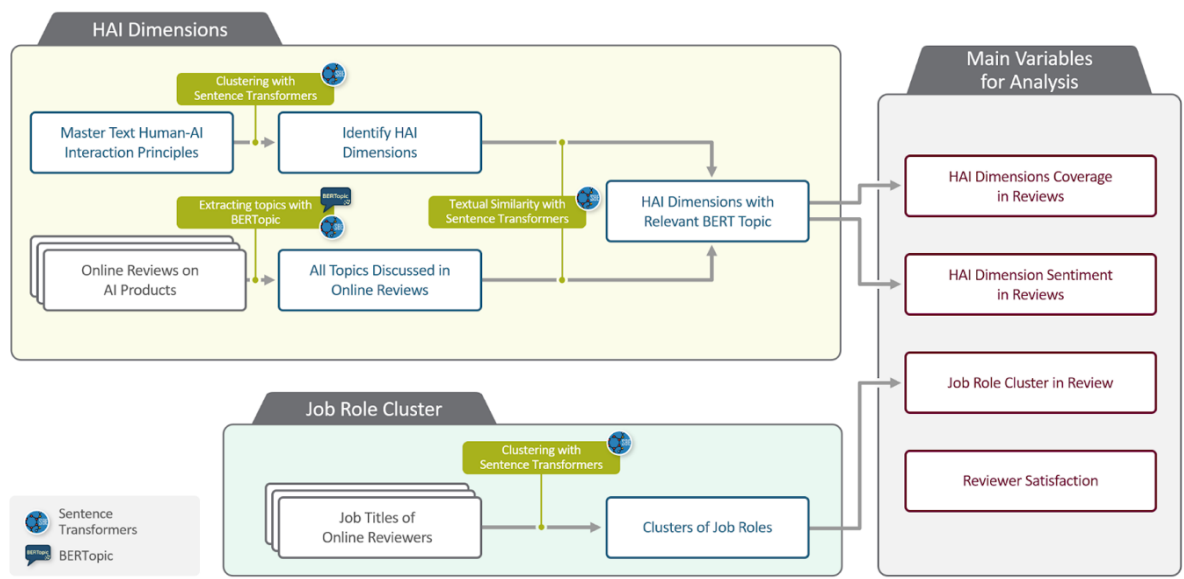

**Figure 1**. Overview of Methodology

This study explores how different Human-AI Interaction (HAI) dimensions are discussed in user reviews and how they relate to user satisfaction. To achieve this, we follow a structured process, as illustrated in Figure 1. We begin by examining industry guidelines for Human-AI Interaction (HAI) and identify key HAI dimensions such as explainability, customization, and adaptability. Using advanced natural language processing (NLP) techniques, we measure how frequently and how positively these aspects are discussed in user reviews.

Beyond the content of reviews, we also analyze who is writing them. By clustering job titles, we identify distinct reviewer groups, allowing us to explore whether different professional backgrounds influence engagement with HAI topics. These measures enable us to assess the relationship and interaction between HAI dimensions, job roles, and user satisfaction, as reflected in review scores.

**Data**

This study relies on online user reviews from G2.com, one of the largest software review platforms. G2 is a widely recognized business-focused review site that specializes in B2B software evaluations, distinguishing it from consumer-oriented platforms like Amazon or Trustpilot, which primarily focus on general consumer products (G2, 2024; Kevans, 2023). This focus on business software products makes G2 particularly well-suited for studying AI-related tools used in professional settings.

We gathered all reviews from G2's "Artificial Intelligence" category, which includes a diverse range of AI-driven products such as Data Science platforms, Chatbots and Voice Assistants, and Generative AI tools. To ensure sufficient data coverage, we included only products with at least 50 reviews, resulting in a dataset of 249 AI products and 108,998 reviews.

A unique feature of G2's review format is that it segments user feedback into structured sections: (i) "What do you like about Product X?" (like section) and "What do you dislike about Product X?" (dislike section). This structured format provides a clear separation between positive and negative sentiments, eliminating the need for complex NLP-based sentiment classification to measure the sentiment of topics discussed (Koch and Pasch, 2023). By analyzing these two sections separately, we can directly assess how different HAI dimensions are perceived—whether positively or negatively.

Additionally, G2 reviews include self-reported job roles of reviewers, enabling an analysis of how different professional backgrounds engage with AI-related topics.

**Measuring HAI Dimensions**

In order to measure concepts related to HAI from free text, the first step in our approach was to gather a comprehensive list of words (master text) related to HAI practices and guidelines. To do so, we drew from three widely recognized industry frameworks and guidelines for HAI: Microsoft's Guidelines for Human-AI Interaction (Microsoft, 2019), Google's People + AI Design Guidebook (Google, 2021), and Apple's Design Guidelines for Machine Learning (Apple, 2023). These frameworks provide structured principles for designing AI systems that are transparent, adaptable, and user-centric. While a broad body of literature has explored various aspects of HAI (for a review, see Raees et al., 2024), we rely on these three guidelines for the following reasons:
- Industry-Focus: Microsoft, Google, and Apple are pioneers in AI and human-computer interaction (HCI), shaping best practices in user experience (UX) design of AI systems for practitioners.

- Comprehensive coverage: Each framework provides structured principles for designing AI interactions that prioritize user needs, ensuring alignment with human-centered AI principles.
- Actionable guidelines and best practices: Unlike purely conceptual frameworks, these guidelines provide practical and applicable design principles that AI developers and designers can implement. This makes them well-suited for evaluating AI products in real-world settings.

Using these frameworks, we extracted key guidelines related to HAI. To ensure comprehensive coverage, we consolidated and harmonized overlapping concepts, removing redundancy while maintaining conceptual clarity. The final set of HAI principles and practices are presented in Table 1, where we indicate which concepts are covered by each framework.

**Table 1: HAI Frameworks and HAI Dimensions**

| Topic | MS Guidelines for HAI | Google People + AI | Apple Guidelines for ML | HAI Dimension | System/ Interaction Focused |
|---|---|---|---|---|---|
| Feedback | ✓ | ✓ | ✓ | 1 | Interaction-focused |
| Accessibility |  | ✓ | ✓ | 2 | Interaction-focused |
| Adaptability | ✓ | ✓ | ✓ | 2 | Interaction-focused |
| Context-awareness | ✓ | ✓ |  | 2 | Interaction-focused |
| Responsiveness | ✓ | ✓ |  | 2 | Interaction-focused |
| Scalability | ✓ | ✓ |  | 2 | Interaction-focused |
| Customization | ✓ | ✓ | ✓ | 3 | Interaction-focused |
| Personalization | ✓ | ✓ |  | 3 | Interaction-focused |
| Ethical AI | ✓ | ✓ |  | 4 | System-focused |
| Explainability | ✓ | ✓ |  | 4 | System-focused |
| Interpretability | ✓ | ✓ |  | 4 | System-focused |
| Learnability | ✓ | ✓ |  | 4 | System-focused |
| Error-recovery | ✓ |  | ✓ | 5 | System-focused |
| Consistency |  | ✓ |  | 6 | System-focused |
| Reliability | ✓ | ✓ | ✓ | 6 | System-focused |
| Trustworthiness | ✓ |  | ✓ | 6 | System-focused |
| Privacy | ✓ | ✓ | ✓ | 7 | System-focused |
| Security | ✓ | ✓ | ✓ | 7 | System-focused |
| Transparency | ✓ | ✓ | ✓ | 7 | System-focused |

After collecting and structuring these HAI concepts, the next step was to identify patterns and relationships among them. To do this, we applied an unsupervised clustering approach to categorize these topics into meaningful dimensions.

First, we transformed each guideline from Table 1 into numerical representations using Sentence Transformers, specifically the model *all-MPNet-base-v2* (Reimers, 2019). Sentence Transformers are advanced language models designed to capture the meaning of text rather than simply matching exact words. Unlike basic word-matching techniques such as keyword searches or word frequency counts, these models recognize deeper semantic relationships. For example, terms like "explainability" and "interpretability" are distinct words, but they are related in meaning. Sentence Transformers allow us to recognize these relationships by positioning similar concepts closer together in a multi-dimensional space, while unrelated terms remain farther apart. This ensures that our clustering approach groups conceptually related topics rather than just those with overlapping wording.

With the vectorized Sentence Transformers representations in place, we employed K-Means clustering to group similar concepts together. To determine the optimal number of clusters, we examined both the Elbow Method and the Silhouette Score. While the Elbow Method provided a more gradual decline without a clear inflection point, the Silhouette Score suggested that a seven-cluster solution best captured the underlying structure of the data (see Appendix A).

The resulting clusters were manually analyzed and labeled based on their shared characteristics. The topic-cluster distribution is represented in Table 1, showing how concepts are grouped together. While the dimensions could naturally be labeled in multiple ways—since each cluster reflects a range of HAI topics—the use of distinct labels serves to enhance interpretability and facilitate clearer discussion in the analysis. The seven identified HAI dimensions are:
- HAI Dimension 1 - Feedback: Emphasizing user feedback mechanisms in AI systems.
- HAI Dimension 2 - Adaptability: AI's responsiveness to user needs, environment, and scalability.
- HAI Dimension 3 - Customization: User-specific adaptations for enhancing AI experience.
- HAI Dimension 4 - Explainability: Concepts related to transparency, ethical considerations, and interpretability of AI models.
- HAI Dimension 5 - Error Recovery: Mechanisms ensuring fault tolerance and efficient handling of AI errors.
- HAI Dimension 6 - Reliability: Ensuring consistency, reliability, and trust in AI interactions.
- HAI Dimension 7 - Privacy & Security: Emphasizing data protection, transparency, and user security in AI applications.

**Categorizing HAI Dimensions: System-Focused vs. Interaction-Focused**

To analyze how different users discuss and value HAI dimensions (Hypotheses H2a and H2b), we classify them into two overarching categories: system-focused and interaction-focused.

This distinction reflects whether a dimension primarily concerns the internal functionality and robustness of AI or the quality of user interaction and adaptability (Amershi et al., 2019; Doshi-Velez & Kim, 2017). Naturally, since we are studying Human-*AI Interaction*, all dimensions inherently relate to both the AI system and its interaction with users to some extent. Nonetheless, they differ in the degree to which they emphasize either the technical system components or the user-facing aspects of interaction and categorizing them by their dominant focus allows for clearer interpretation and analysis.

**Interaction-Focused HAI Dimensions**: These dimensions relate to how AI systems engage with users, ensuring adaptability, responsiveness, and personalization in interactions.
- HAI Dimension 1 – Feedback: AI systems often incorporate user feedback mechanisms to refine and improve performance. This dimension reflects the extent to which AI can learn from user input and adjust interactions accordingly, making the system feel more dynamic and responsive (Eiband et al., 2018). Since feedback loops primarily enhance user engagement and interaction, this dimension is classified as interaction-focused.
- HAI Dimension 2 – Adaptability: AI's ability to adjust to user needs and environmental factors is crucial for seamless interaction. This includes context-aware recommendations, dynamic responses, and adaptive learning capabilities (Yang et al., 2020). Because this dimension directly affects how intuitive and user-centric AI feels, it aligns with interaction-focused HAI principles.
- HAI Dimension 3 – Customization: With customization and personalization, AI technologies aim to tailor experiences to individual users (Knijnenburg et al., 2012). Personalization includes adjustable settings, adaptive recommendations, and flexible AI behavior, all of which enhance user interaction rather than system integrity. Thus, this dimension is classified as interaction-focused.

**System-Focused HAI Dimensions:** These dimensions concern AI's internal reliability, transparency, and robustness, ensuring predictable and responsible system behavior.
- HAI Dimension 4 – Explainability: This dimension encompasses the principles of AI transparency, interpretability, and ethical alignment (Lipton, 2016; Doshi-Velez & Kim, 2017). Explainability is critical for users who need to understand AI decision-making, and ethical AI ensures that systems function in a socially responsible manner. Since both aspects concern the internal logic and governance of AI, this dimension is categorized as system-focused.
- HAI Dimension 5 – Error Recovery: AI's ability to handle failures, self-correct, and minimize disruption is crucial for system functioning and long-term trust (Corso et al., 2023). Effective error recovery mechanisms ensure system robustness rather than shaping direct user interactions, making this a system-focused dimension.
- HAI Dimension 6 – Reliability: This dimension encompasses reliability, consistency, and trustworthiness, all of which are fundamental to ensuring that AI systems function dependably and predictably (Xiong et al., 2022). A reliable and consistent AI system delivers accurate and stable results, minimizing unexpected failures and maintaining operational integrity. While trustworthiness can also be shaped by user-facing factors such as personalization or feedback, this dimension emphasizes trust grounded in technical robustness. Since its core concern is the dependable operation of the underlying system, this dimension is classified as system-focused.
- HAI Dimension 7 – Privacy & Security: Protecting user data, ensuring secure AI operations, and complying with regulatory standards are central to AI's technical integrity (Usmani et al., 2023). Since these concerns primarily involve system architecture and governance, this dimension is classified as system-focused.

**Topic Modeling of Online Reviews**

To measure HAI dimensions in user discussions, we first identified all topics mentioned in online reviews of AI products—whether related to Human-AI Interaction or broader product experiences. For this, we employed BERTopic (Grootendorst, 2022), an advanced topic modeling technique that automatically identifies topics within a dataset. Unlike traditional clustering methods such as K-Means or Latent Dirichlet Allocation (LDA) (Blei et al., 2003), which require a pre-defined number of clusters, BERTopic dynamically structures topics based on the data itself, making it particularly well-suited for capturing diverse themes in user-generated content. A key advantage of BERTopic is that it also relies on Sentence Transformer embeddings to encode the semantic meaning of the text data. This allows the model to identify topics based on meaning rather than simple word frequency, capturing relationships between conceptually similar terms even when they do not share exact wording.

Moreover, BERTopic assigns probability scores to multiple topics within each review rather than forcing each text into a single category ("soft clustering"). Traditional topic models, such as LDA, assume that each document belongs to a single dominant topic. However, AI product reviews often discuss multiple themes simultaneously—for example, a review may highlight an AI tool's adaptability while also mentioning concerns about its security. Similarly, user reviews frequently blend broader product-related aspects, such as pricing or general product category descriptions. By capturing the degree to which multiple topics are present in a text, BERTopic helps disentangle discussions of HAI concepts from other themes.

Notably, for this topic modeling analysis, we measured the topics separately in the "like" and "dislike" sections. By doing so, we can disentangle which topics are discussed in a positive and which in a negative way, allowing us to also measure sentiment on HAI discussions.

**Identifying Topics on HAI**

To extract topics related to the HAI dimensions, we leveraged an embedding-based similarity approach. First, we used Sentence Transformers to encode the top 10 words associated with each topic identified by BERTopic to get a representation of each topic. In parallel, we embedded the seven Human-Centered AI clusters using their associated principles from Table 1, which were also vectorized using Sentence Transformers.

Next, we measured the textual similarity between all BERTopic topics and the seven HAI dimensions. For each dimension, we identified the 10 most similar topics by computing cosine similarity scores between the topic embeddings and the embeddings (Reimers, 2019). By doing so, we mapped the most related topics from BERTopic to the seven HAI dimensions, ensuring that only the most semantically relevant topics were assigned based on their similarity scores. Naturally, there are multiple ways to map BERTopic-derived topics to the seven HAI dimensions. A key decision involves determining how many topics should be assigned to each dimension. One option would have been to include a larger number of topics per dimension (e.g., 20 or 30 topics). However, we found that increasing the number of assigned topics often introduced topics that were only loosely related to HAI dimensions. By restricting the selection to the top 10 most similar topics per dimension, we ensured that the mapped topics were strongly relevant while still capturing meaningful variation in user discussions.

Another approach could have been to set a fixed similarity score threshold for assigning topics. However, this would have led to an inconsistent number of topics per dimension, making comparisons across HAI dimensions difficult. Some dimensions might have received a large number of assigned topics, while others with slightly lower similarity scores might have had far fewer, creating imbalances in the analysis.

By selecting the top 10 most similar topics per HAI dimension, we balanced the trade-off between coverage and interpretability, ensuring a structured and meaningful mapping while avoiding excessive noise or topic dilution. The results of this mapping are presented in Appendix B, where each HAI dimension is associated with its most relevant topics.

**Measurement of Human-AI Interaction Dimensions**

To measure how Human-AI Interaction (HAI) dimensions are discussed in user reviews, we compute two key metrics for each review:
- **Sentiment Score** – This measures whether a given HAI dimension is discussed positively or negatively. If a dimension appears more frequently in the "like" section, it receives a positive sentiment score, whereas if it is mentioned more often in the "dislike" section, it receives a negative score. A value close to zero suggests neutral discussion or minimal mention of the dimension.
- **Coverage Score** – This quantifies how strongly a given HAI dimension is discussed in a review, regardless of sentiment. A higher score indicates that the dimension is a key topic in the review, while a lower score suggests it is not or only briefly mentioned.

To compute these scores, we use BERTopic's probabilistic modeling, which assigns probability values to topics rather than strictly categorizing each review. Since each HAI dimension consists of multiple related topics, we sum the probability scores of all topics within a given dimension to derive its overall Sentiment and Coverage scores for a review. The Sentiment Score for an HAI dimension is calculated by summing the probabilities of all related topics appearing in the "like" section and subtracting the probabilities of those topics in the "dislike" section. This ensures that dimensions frequently mentioned positively receive higher scores, while those discussed negatively have lower (negative) scores. The Coverage Score, on the other hand, is determined by summing the probabilities of all related topics across both the

"like" and "dislike" sections. This captures the overall presence of a dimension in the review without considering whether it is framed positively or negatively. By aggregating probabilities across multiple topics per dimension, we ensure that our measurement reflects the overall emphasis placed on each HAI dimension in user discussions.

**Clustering Job Roles**

To better understand how different user groups engage with HAI concepts, we also clustered the job titles of reviewers from the G2 dataset. Again, we first embedded all job title names using Sentence Transformer (*'all-MPNet-base-v2'*). We then applied K-Means clustering to group similar job titles together. To determine the optimal number of clusters, we used both the Elbow Method and Silhouette Score, identifying five clusters for job titles. Although the global maximum Silhouette Score occurred at k = 2, this solution proved too coarse: for example, it grouped software engineers with administrative assistants and sales representatives, failing to reflect meaningful differences in technical expertise. Instead, we utilized the local maximum at k = 5, which offered a stable clustering solution that balanced interpretability with subgroup distinction (see Appendix A)

The final clustering yielded five broad professional categories:
- Job Cluster 1: Writers & Authors (e.g., bloggers, editors, translators)
- Job Cluster 2: Marketing & Sales Professionals (e.g., marketing managers, product managers, sales consultants)
- Job Cluster 3: Business & Operations Roles (e.g., project managers, consultants, operations manager)
- Job Cluster 4: Executives & Founders (e.g., CEOs, founders, directors, managing partners)
- Job Cluster 5: Technical Roles (e.g., software engineers, data scientists, developers)

Importantly, Job Cluster 5 closely aligns with the technical user group described in Hypothesis 2—professionals involved in developing, implementing, and maintaining technology. The remaining clusters represent a diverse set of non-technical roles or domain specialists, offering a broader view of how AI is applied in everyday professional tasks beyond system development. Appendix C shows the top 10 most frequently occurring job titles within each cluster.

## 4. Results

**HAI Sentiment and Overall Satisfaction**

To analyze how sentiment toward HAI dimensions relates to user satisfaction, we regressed the sentiment score of each dimension on the reviewers' overall rating (1-5 scale). The sentiment score for a cluster measures the extent to which a user discusses a topic in the "like" section while not mentioning it in the "dislike" section. A score of 1 means that the user exclusively discusses topics on a dimension positively without expressing any negative remarks. Table 2 presents the regression results. The first model ("HAI Dim. Only") includes only the measures of sentiment on the seven HAI dimensions, while the second model ("With Controls") incorporates product category, company age, and company size (number of employees) to account for potential confounding effects.

Our analysis reveals that not all Human-AI Interaction dimensions are equally associated with user satisfaction. Among the seven identified dimensions, we find positive and significant effects for adaptability, customization, error recovery, and privacy & security, while feedback, explainability, and reliability showed no significant effects.

Among these, error recovery has the strongest effect on user satisfaction. Reviews that only discuss error recovery in a fully positive way see a 0.21 to 0.25-point increase in review scores. This corresponds to an increase of roughly 0.35 standard deviations, indicating that users place high value on AI systems that can effectively handle errors and failures. Similarly, positive discussions of privacy-related features are associated with an increase of roughly 0.1 points in review scores.

Among the interaction-focused dimensions, we find positive and significant effects for adaptability (HAI Dimension 2) and customization (HAI Dimension 3), with effect sizes of 0.18 and 0.11 points, respectively. These findings suggest that users favor AI products that are flexible, responsive to their needs, and capable of personalization.

Overall, these findings provide partial support for H1, as sentiment toward four out of the seven HAI dimensions is positively associated with user satisfaction.

**Table 2: Human-Centered AI Sentiment and User Satisfaction**
Dependent Variable: User Satisfaction

| Variable | HAI Dim. Only | With Controls |
|---|---|---|
| HAI D1- Feedback | 0.013 | -0.002 |
|  | (0.034) | (0.034) |
| HAI D2 - Adaptability | 0.180** | 0.176* |
|  | (0.085) | (0.102) |
| HAI D3 - Customization | 0.119** | 0.114** |
|  | (0.052) | (0.049) |
| HAI D4 - Explainability | 0.038 | 0.046 |
|  | (0.053) | (0.063) |
| HAI D5 - Error Recovery | 0.254*** | 0.215*** |
|  | (0.068) | (0.071) |
| HAI D6 - Reliability | 0.026 | 0.079 |
|  | (0.070) | (0.060) |
| HAI D7 - Privacy | 0.115* | 0.099* |
|  | (0.059) | (0.052) |
| Observations | 108998 | 103361 |
| Controls | No | Yes |
| R-squared | 0.00 | 0.02 |

Robust Standard errors in parentheses.* p<.1, ** p<.05, ***p<.01. Controls include: Dummies for product category, company age, and number of employees.

## Professional Background and the Coverage of HAI Dimensions

This section examines how a reviewer's job role influences the likelihood of discussing specific HAI dimensions in their reviews. Table 3 presents the coverage of HAI dimensions as the dependent variable, capturing how strongly corresponding topics are discussed—regardless of sentiment. For easier interpretation, the dependent variables are standardized with Z-scores. The key independent variables are indicator variables for different job role clusters, with Cluster 1 ("writers") serving as the reference category.

Overall, the results provide support for Hypothesis H2a. Reviewers in technical roles (Job Cluster 5) discuss two out of three interaction-focused HAI dimensions significantly less often than non-technical reviewers. Specifically, they mention HAI Dimension 1 (feedback) 0.05 standard deviations less and HAI Dimension 3 (customization) 0.03 standard deviations less than writers. We find no significant difference for HAI Dimension 2 (adaptability)—potentially because this dimension also includes elements relevant to system functionality (e.g., scalability).

Conversely, three out of four system-focused HAI dimensions (explainability, error recovery, and reliability) are significantly more discussed by reviewers in technical roles. The strongest effect is observed for HAI Dimension 4 (explainability), which is covered 0.06 standard deviations more often by technical professionals than by non-technical reviewers. Smaller but still significant effects are found for HAI Dimension 5 (error recovery) and HAI Dimension 6 (reliability), both of which are discussed 0.02 standard deviations more often by technical professionals. The effect of technical job roles on HAI Dimension 7 (privacy & security) is positive but non-significant, suggesting that concerns on these topics are relevant to both technical and non-technical users.

Table 3: Coverage of HAI Dimensions and Job Roles
Dependent Variable: Coverage of HAI Dimensions

| Variable | Interaction-Focused | | | System-Focused | | | |
|---|---|---|---|---|---|---|---|
| | HAI D1 Feedback | HAI D2 Adapt. | HAI D3 Custom. | HAI D4 Explain. | HAI D5 Error Rec. | HAI D6 Reliability | HAI D7 Privacy |
| Job Cluster 2 *Marketing* | 0.031*** (0.010) | -0.009 (0.010) | 0.001 (0.010) | -0.012 (0.010) | -0.016 (0.010) | -0.022** (0.010) | -0.012 (0.010) |
| Job Cluster 3 *Business* | 0.034*** (0.010) | -0.011 (0.010) | -0.017 (0.010) | -0.005 (0.010) | -0.004 (0.010) | -0.003 (0.010) | -0.008 (0.010) |
| Job Cluster 4 *Executives* | 0.006 (0.010) | -0.014 (0.011) | -0.013 (0.011) | -0.008 (0.011) | -0.005 (0.011) | -0.010 (0.011) | 0.010 (0.011) |
| Job Cluster 5 *Technical* | -0.052*** (0.011) | 0.011 (0.011) | -0.037*** (0.011) | 0.064*** (0.011) | 0.024** (0.011) | 0.021* (0.011) | 0.010 (0.011) |
| Observations | 103361 | 103361 | 103361 | 103361 | 103361 | 103361 | 103361 |
| Controls | Yes | Yes | Yes | Yes | Yes | Yes | Yes |
| R-squared | 0.03 | 0.00 | 0.01 | 0.00 | 0.00 | 0.00 | 0.01 |

Robust Standard errors in parentheses.* $p<.1$, ** $p<.05$, ***$p<.01$. Controls include: Dummies for product category, company age, and number of employees

Among non-technical job role clusters, differences in HAI topic coverage are minimal. A significant difference, however, is found for HAI Dimension 1 (feedback), which is discussed more frequently by marketing (Job Cluster 2) and business professionals (Job Cluster 3) compared to writers. This is unsurprising given their client-facing roles and focus on customer interactions. In contrast, marketing professionals mention HAI Dimension 6 (reliability) significantly less often than writers.

**Job Roles as a Moderator of HAI Sentiment on User Satisfaction**

In this section, we examine whether job roles influence the relationship between HAI sentiment and user satisfaction, with a particular focus on the effect of technical roles (Hypothesis H2b). To test this, we estimated interaction effects between HAI sentiment and job clusters in predicting satisfaction scores. Given the large number of interaction terms—35 (7 HAI dimensions × 5 job clusters)—the results are summarized in Figure 2.

The findings indicate that job roles do not meaningfully moderate the relationship between HAI sentiment and satisfaction. Only 2 out of 35 interaction terms reach statistical significance at the p < 0.05 level. However, when applying a Bonferroni correction (Bonferroni, 1936) for multiple comparisons, only one interaction remains significant—between HAI Dimension 4 (explainability) and Job Cluster 2 (marketing & sales). This suggests that, overall, job roles do not substantially alter the effect of HAI sentiment on satisfaction.

Crucially, we find no significant interactions for technical roles (Job Cluster 5), meaning that the effect of HAI sentiment on satisfaction does not differ between technical and non-technical users. This means that we find no support for Hypothesis H2b.

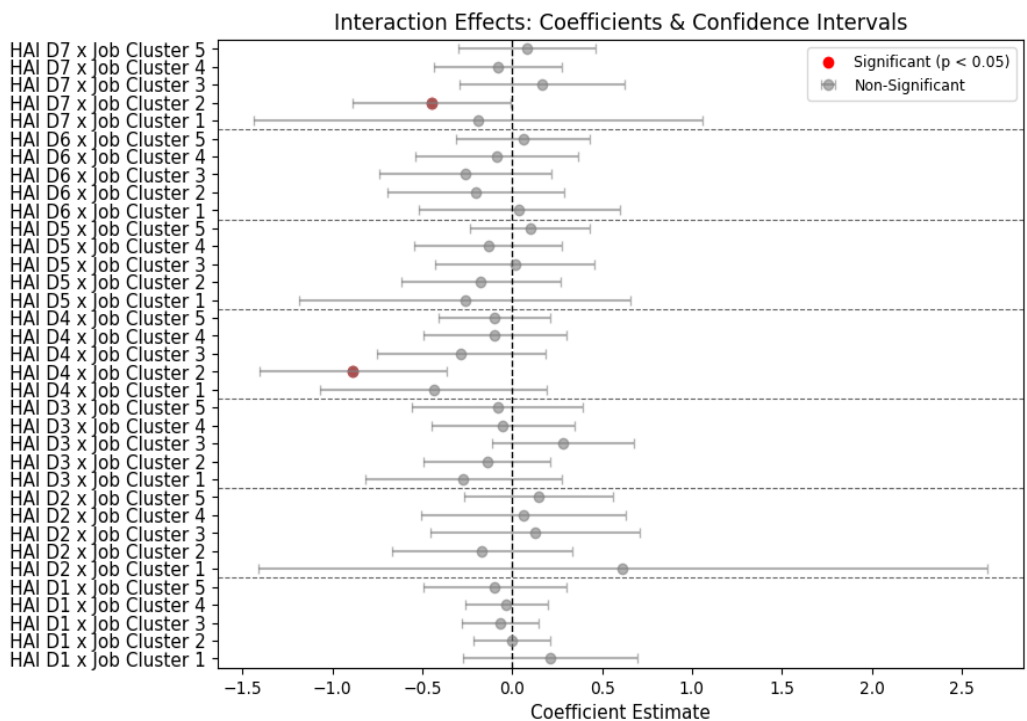

**Figure 2**: Effect on overall satisfaction. Interaction effects between HAI dimensions and Job Roles

# 5. Discussion:

The findings of this study contribute to our understanding of how Human-AI Interaction (HAI) concepts influence user satisfaction, offering large-scale empirical evidence that connects HAI dimensions with user evaluations in real-world AI product reviews. Beyond review content, we also account for user heterogeneity by incorporating job roles, shedding light on how different types of users engage with and evaluate HAI practices.

**The Effect of HAI Principles on User Satisfaction**

In line with Hypothesis H1, we find (partial) support for the idea that user sentiment toward HAI dimensions is positively associated with satisfaction. Specifically, four out of the seven dimensions show a statistically significant relationship with review scores. This suggests that when users perceive aspects emphasized in HAI guidelines positively—indicating that those HAI design principles are successfully implemented in the product—their satisfaction with the overall AI product increases accordingly.

This finding is important as it provides empirical support for the practical value of HAI guidelines, which are often grounded in expert consensus or qualitative evidence. While prior literature has strongly advocated for human-centered AI design (e.g., Amershi et al., 2019; Shneiderman, 2020), large-scale, user-based validation has been limited. Our study addresses this gap by quantitatively linking user perceptions of HAI principles to their satisfaction in a diverse set of AI products.

Notably, we observe significant effects for both system-focused (error recovery and privacy & security) and interaction-focused (customization and adaptability) HAI dimensions. This aligns with prior research (e.g., Amershi et al., 2019), which emphasizes that HAI guidelines should not be limited to technical performance but should also address the full workflow between users and AI. The positive effects of customization and adaptability highlight the importance of AI systems that can flexibly respond to user needs, while the strong association between error recovery and satisfaction underscores the critical role of AI systems in managing failures effectively. The significant effect of privacy & security suggests that concerns over data protection and safety meaningfully shape user perceptions.

While three HAI dimensions—feedback, explainability, and reliability—do not show a significant effect, this does not necessarily imply they are unimportant. Some principles, such as explainability and reliability, may function as baseline expectations—users only notice them when they are absent or fail. Others may be context-dependent, playing a more critical role in specific AI applications (e.g., finance, healthcare) rather than in general-use AI tools. Additionally, measurement constraints may contribute to the lack of significance, as users may appreciate these principles without explicitly referencing them in reviews.

Overall, our findings support the idea that sentiment toward well-implemented HAI principles correlates with user satisfaction. However, the effects are not uniform across all dimensions, highlighting the need for nuanced design strategies that prioritize user-perceived value.

**The Role of Professional Background on Human-AI Interaction**

To further understand how different users engage with Human-AI Interaction (HAI) principles, we examined the role of professional background by clustering reviewers based on their job titles. Our analysis reveals meaningful differences in the coverage of HAI dimensions across job roles, lending support to Hypothesis H2a.

In particular, reviewers with technical job roles—such as engineers, developers, and data scientists—are significantly more likely to discuss system-focused HAI dimensions. These dimensions correspond closely to how AI systems function internally and are therefore particularly salient to users who interact with or maintain system infrastructure. Conversely, these same reviewers are significantly less likely to mention interaction-focused dimensions, such as feedback and customization, which are more directly related to front-end user experience. Interestingly, there was no significant difference for adaptability, potentially due to its hybrid nature—capturing both user-facing flexibility and system-level scalability.

These findings complement earlier research in Human-Computer Interaction, which has long emphasized that user roles shape how technologies are evaluated and experienced (Grudin, 1991; Ko et al., 2006). By linking professional background to the salience of HAI dimensions in large-scale, user-generated review data, this study offers empirical validation of role-based differences specifically within the domain of AI interaction. In doing so, it extends existing HCI theory into the context of HAI and provides quantitative evidence for the distinction between system- and interaction-focused concerns in real-world evaluations of AI systems.

Among non-technical job clusters, such as writers and marketing professionals, we observe only minimal differences in the discussion of HAI dimensions. This suggests that HAI concepts like customization, privacy, or feedback are equally relevant across a range of non-technical use cases—such as content creation, marketing, sales, or consultanting—highlighting their broad importance in applied, day-to-day AI use.

However, when testing whether job roles moderate the effect of HAI sentiment on satisfaction (Hypothesis H2b), we find no support for this idea. While technical users are more likely to discuss certain HAI dimensions, this does not influence how strongly sentiment on those dimensions is associated with overall product satisfaction. Out of 35 tested interaction effects, only one was significant after correcting for multiple comparisons. This suggests a universal effect of HAI dimensions: once a specific HAI dimension is salient to a user—regardless of their professional background—its impact on satisfaction is consistent.

Taken together, these findings offer a nuanced perspective. While professional background shapes which aspects of HAI users focus on, it does not significantly alter how these perceptions translate into satisfaction. This points to the generalizability of HAI principles across user roles and the broad relevance of both system- and interaction-focused dimensions in shaping the user experience with AI products.

# 6. Conclusion

**Implications**

The results of this study have several important implications for the design, evaluation, and deployment of AI systems in practice. Our findings underscore that certain Human-AI Interaction (HAI) dimensions—particularly adaptability, customization, error recovery, and privacy—are closely linked to user satisfaction. This suggests that system designers should prioritize these dimensions not just from an ethical or usability standpoint, but as key levers for improving overall user experience and acceptance of AI technologies. By aligning product development with user-perceived strengths in these areas, organizations can foster more positive user outcomes and potentially drive adoption.

The analysis of job roles reveals that different users bring different priorities to their evaluation of AI systems. Technical professionals emphasize system-level capabilities like explainability and reliability, while non-technical users focus more on usability and personalization. This highlights the need for role-sensitive AI design—one that supports both deep system transparency for expert users and intuitive, user-friendly interfaces for general users. Adaptive interfaces that adjust based on user background or preferences may help reconcile these differing needs.

While user roles shape which dimensions are discussed, our findings suggest that once a specific HAI aspect is noticed and evaluated positively or negatively, its effect on satisfaction is relatively universal. This implies that HAI principles are not niche concerns, but broadly relevant features that matter to a wide range of users. For practitioners, this supports the value of implementing core HAI principles universally—while still designing interfaces and features that can flexibly adapt to different user backgrounds and needs.

**Limitations and Future Research**

This study provides large-scale empirical evidence on how Human-AI Interaction (HAI) dimensions relate to user satisfaction using user-generated review data. However, several limitations should be acknowledged, which also offer valuable directions for future research.

First, our analysis is based on a specific set of HAI guidelines derived from industry leaders—Microsoft, Google, and Apple. While these frameworks offer practical and widely adopted design recommendations, a broad body of literature has explored HAI from academic, ethical, and domain-specific perspectives (for a review, see Raees et al., 2024). Given the methodological complexity of our approach a comparison of multiple frameworks was beyond the scope of this study. Future research could evaluate how findings vary when using alternative frameworks or guidelines of HAI.

Second, although we control for product category, our main analysis does not examine in depth how different types of AI products—such as chatbots versus data science platforms—shape the relevance or salience of specific HAI dimensions. While an in-depth examination and discussion of these differences was beyond the scope of this study, future research could explore how product functionality, domain, or user context interact with HAI concerns.

Third, while we measure *sentiment* on HAI dimensions—that is, whether users evaluate a given dimension positively or negatively—we do not examine the underlying *reasons* behind these evaluations. Future research could explore what specific features or experiences contribute to satisfaction or dissatisfaction with particular HAI principles, potentially using qualitative or experimental methods.

Finally, while BERTopic offers several advantages—including contextual embeddings and multi-topic assignment—it remains an unsupervised method that involves interpretive steps when grouping topics into broader HAI dimensions. Although we mitigate this through systematic mapping, results may still vary depending on modeling parameters and judgment calls. Future work could combine this approach with supervised learning, expert coding, or human-in-the-loop strategies to further validate and refine topic-to-dimension mappings.

# References

Apple (2023). Available: Human Interface Guidelines for Machine Learning Available: https://developer.apple.com/design/human-interface-guidelines/machine-learning

Amershi, S., Cakmak, M., Knox, W. B., & Kulesza, T. (2014). Power to the people: The role of humans in interactive machine learning. *AI magazine*, *35*(4), 105-120.

Amershi, S., Weld, D., Vorvoreanu, M., Fourney, A., Nushi, B., Collisson, P., ... & Horvitz, E. (2019, May). Guidelines for human-AI interaction. In *Proceedings of the 2019 chi conference on human factors in computing systems* (pp. 1-13).

Blei, D. M., Ng, A. Y., & Jordan, M. I. (2003). Latent Dirichlet Allocation. *Journal of Machine Learning Research, 3*, 993–1022.

Blackwell, A. F. (2002, September). First steps in programming: A rationale for attention investment models. In *Proceedings IEEE 2002 Symposia on Human Centric Computing Languages and Environments* (pp. 2-10). IEEE.

Bonferroni, C. (1936). Teoria statistica delle classi e calcolo delle probabilita. *Pubblicazioni del R istituto superiore di scienze economiche e commericiali di firenze*, *8*, 3-62.

Carroll, J. M. (1997). Human–computer interaction: Psychology as a science of design. *International journal of human-computer studies*, *46*(4), 501-522.

Corso, A., Karamadian, D., Valentin, R., Cooper, M., & Kochenderfer, M. J. (2023). A holistic assessment of the reliability of machine learning systems. *arXiv preprint arXiv:2307.10586*.

Davis, F. D. (1989). Perceived usefulness, perceived ease of use, and user acceptance of information technology. *MIS quarterly*, 319-340.

Díaz, A., García, A., & Gervás, P. (2008). User-centred versus system-centred evaluation of a personalization system. *Information Processing & Management*, *44*(3), 1293-1307.

Dillon, A., & Watson, C. (1996). User analysis in HCI—the historical lessons from individual differences research. *International journal of human-computer studies*, *45*(6), 619-637.

Eiband, M., Schneider, H., Bilandzic, M., Fazekas-Con, J., Haug, M., & Hussmann, H. (2018, March). Bringing transparency design into practice. In *Proceedings of the 23rd international conference on intelligent user interfaces* (pp. 211-223).

Fischer, G. (2001). User modeling in human–computer interaction. *User modeling and user-adapted interaction*, *11*, 65-86.

G2. (2024). Where you go for software. Available: https://company.g2.com/about
Gasson, S. (2003). Human-centered vs. user-centered approaches to information system design. *Journal of Information Technology Theory and Application (JITTA)*, *5*(2), 5.
Google (2021). Google: People+AI Design Principles: Available: https://pair.withgoogle.com/guidebook/patterns

Grudin, J. (1991). Systematic sources of suboptimal interface design in large product development organizations. *Human-computer interaction*, *6*(2), 147-196.

Grootendorst, M. (2022). BERTopic: Neural topic modeling with a class-based TF-IDF procedure. *arXiv preprint arXiv:2203.05794*.

Kevans, J. (2024). 10 Best Sooftware Review Sites. Available: https://b2bsaasreviews.com/best-software-review-sites

Knijnenburg, B. P., Willemsen, M. C., Gantner, Z., Soncu, H., & Newell, C. (2012). Explaining the user experience of recommender systems. *User modeling and user-adapted interaction*, *22*, 441-504.

Koch, S., & Pasch, S. (2023, December). CultureBERT: Measuring Corporate Culture With Transformer-Based Language Models. In *2023 IEEE International Conference on Big Data (BigData)* (pp. 3176-3184). IEEE.

Ko, A. J., Myers, B. A., Coblenz, M. J., & Aung, H. H. (2006). An exploratory study of how developers seek, relate, and collect relevant information during software maintenance tasks. *IEEE Transactions on software engineering*, *32*(12), 971-987.

Lipton, Z. C. (2018). The mythos of model interpretability: In machine learning, the concept of interpretability is both important and slippery. *Queue*, *16*(3), 31-57.

Pasch, S. (2025). LLM Content Moderation and User Satisfaction: Evidence from Response Refusals in Chatbot Arena. *arXiv preprint arXiv:2501.03266*.

Raees, M., Meijerink, I., Lykourentzou, I., Khan, V. J., & Papangelis, K. (2024). From explainable to interactive AI: A literature review on current trends in human-AI interaction. *International Journal of Human-Computer Studies*, 103301.

Rai, A. (2020). Explainable AI: From black box to glass box. *Journal of the academy of marketing science*, *48*, 137-141.

Reimers, N. (2019). Sentence-BERT: Sentence Embeddings using Siamese BERT-Networks. *arXiv preprint arXiv:1908.10084*.

Shneiderman, B. (2020). Human-centered artificial intelligence: Reliable, safe & trustworthy. *International Journal of Human–Computer Interaction*, *36*(6), 495-504.

Shin, D. (2020). User perceptions of algorithmic decisions in the personalized AI system: Perceptual evaluation of fairness, accountability, transparency, and explainability. *Journal of Broadcasting & Electronic Media*, *64*(4), 541-565.

Sundar, A., Russell-Rose, T., Kruschwitz, U., & Machleit, K. (2024). The AI Interface: Designing for the Ideal Machine-Human Experience. *Computers in Human Behavior*, 108539.

Usmani, U. A., Happonen, A., & Watada, J. (2023, June). Human-centered artificial intelligence: Designing for user empowerment and ethical considerations. In *2023 5th international congress on human-computer interaction, optimization and robotic applications (HORA)* (pp. 1-7). IEEE.

Venkatesh, V., & Davis, F. D. (2000). A theoretical extension of the technology acceptance model: Four longitudinal field studies. *Management science*, *46*(2), 186-204.

Xiong, P., Buffett, S., Iqbal, S., Lamontagne, P., Mamun, M., & Molyneaux, H. (2022). Towards a robust and trustworthy machine learning system development: An engineering perspective. *Journal of Information Security and Applications*, *65*, 103121.

Yang, Q., Steinfeld, A., Rosé, C., & Zimmerman, J. (2020, April). Re-examining whether, why, and how human-AI interaction is uniquely difficult to design. In *Proceedings of the 2020 chi conference on human factors in computing systems* (pp. 1-13).

Yang, Q., & Lee, Y. C. (2024). Ethical AI in financial inclusion: The role of algorithmic fairness on user satisfaction and recommendation. *Big Data and Cognitive Computing*, *8*(9), 105.

Yeomans, M., Shah, A., Mullainathan, S., & Kleinberg, J. (2019). Making sense of recommendations. *Journal of Behavioral Decision Making*, *32*(4), 403-414.

# Appendix

## Appendix A: Clustering

**Clustering of HAI Topics**

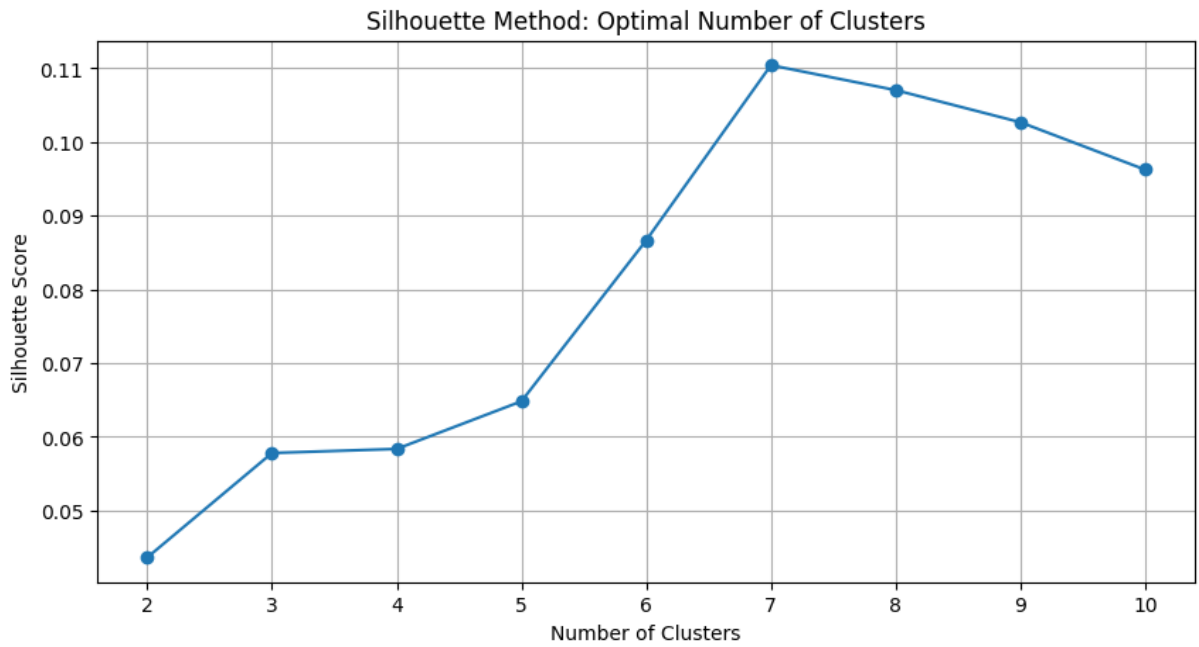

**Figure AF1**: Silhoutte Scores for clustering HAI related topics in different numbers of clusters

**Clustering of Job Roles**

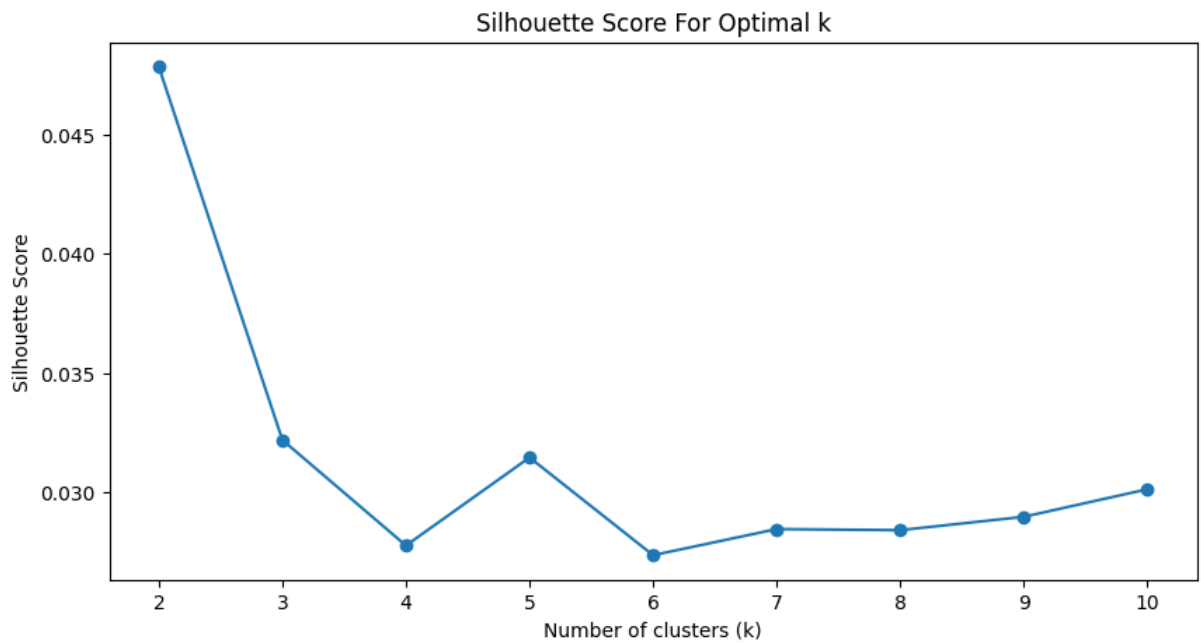

**Figure AF2**: Silhoutte Scores for clustering job titles from online reviews in different numbers of clusters

**Appendix B: HAI Dimensions and Associated Topics from BERTopic**

**HAI Dimension 1: Feedback**
- Topic 552: feedback, mechanism, positive, continuous, feedbacks, elaborate, hearing
- Topic 11: reviews, review, leave, request, send, respond, customers, rating
- Topic 577: negative, feedback, humanly, negatively, feedbacks, personally, positive, NPS
- Topic 125: complaints, complaint, complain, far, honestly, pleasant, satisfied
- Topic 198: survey, surveys, questionnaires, respondents, rating, feedback
- Topic 22: reviews, customers, leave, respond
- Topic 837: reports, reporting, dislikes, centered, dislike
- Topic 27: reporting, reports, customizable, customization
- Topic 846: complaints, complain, satisfying, negatives, flaws
- Topic 917: complain, complaints, enjoyment, program

**HAI Dimension 2: Adaptability**
- Topic 267: accessibility, impaired users, disabilities, navigation
- Topic 850: agility, humanizing technology, performance
- Topic 1043: inventiveness, technology, content adaptability
- Topic 976: responsiveness, caring, eagerness
- Topic 465: centralized knowledge, privacy, efficiency
- Topic 693: usability, technical proficiency, experience
- Topic 878: NLP, automation, user-friendliness
- Topic 281: responsiveness, selection, team performance
- Topic 1184: quality indicators, smarter exposure, adaptability
- Topic 506: interface design, reliability, pain points

**HAI Dimension 3: Customization**
- Topic 593: personalization, personalizer, human-like experiences
- Topic 113: customization, customizability, configurable settings
- Topic 393: widgets, colors, appearance customization
- Topic 27: reporting, customizable reports
- Topic 71: email templates, customization, design flexibility
- Topic 329: settings, preferences, configurations
- Topic 777: customizations, personalization, adaptable options
- Topic 1173: hyper-personalization, contextualized experiences
- Topic 105: layout, formatting preferences
- Topic 227: personalized outreach, AI-driven customization

**HAI Dimension 4: Explainability**
- Topic 878: NLP, automation, tokenization, intelligence, user-friendly, efficiency
- Topic 184: intuitive, intuitiveness, search, modernization
- Topic 24: AI, artificial intelligence, models, machine learning
- Topic 1007: probabilistic reasoning, business-focused, modeling, inference, algorithms
- Topic 512: comprehension, sentiment analysis, extracted, simplification
- Topic 452: conversational AI, chatbots, dialogue
- Topic 504: rule-based systems, declarative logic, engine
- Topic 1011: observability, monitoring, profiling, ML models
- Topic 92: NLP, natural language processing, entity recognition, sentiment
- Topic 527: BERT, embeddings, biases, predictions

**HAI Dimension 5: Error Recovery**

- Topic 869: debugging, errors, issue resolution
- Topic 1133: crashes, debugging, troubleshooting
- Topic 457: outages, downtime, service recovery
- Topic 455: issue resolution, service continuity
- Topic 428: autosave, error prevention, undo features
- Topic 672: cluster failures, job restarts, retry mechanisms
- Topic 496: glitches, app stability, user experience
- Topic 274: error codes, debugging messages
- Topic 47: installation, configuration, setup
- Topic 788: debugging tools, issue characterization

### HAI Dimension 6: Reliability & Trustworthiness
- Topic 1184: quality indicators, reliability, robustness
- Topic 692: stability, system performance, integration
- Topic 850: agility, corporate solutions, efficiency
- Topic 185: screen sharing, remote communication
- Topic 536: communication clarity, issue resolution
- Topic 627: partnerships, reliability in service delivery
- Topic 353: flexibility, consistency, product reliability
- Topic 693: efficient user experience, usability
- Topic 1028: trustworthiness, product legitimacy
- Topic 435: equipment reliability, security compliance

### HAI Dimension 7: Privacy & Security
- Topic 333: privacy, security, data protection
- Topic 304: cyber security, hacking threats, secure access
- Topic 803: browsing privacy, online security
- Topic 465: centralized data privacy, operational security
- Topic 1154: GDPR compliance, regulatory adherence
- Topic 78: surveillance, camera security, motion detection
- Topic 910: transparency, visibility, information access
- Topic 920: personal data management, information control
- Topic 345: security monitoring, incident reporting
- Topic 80: knowledge security, access control

**Appendix C: Job Titles by Job Cluster**

**Table AF1: Most Mentioned Job Titles by Job Cluster**

| Job Cluster | Top 10 Job Titles |
|---|---|
| Cluster 1: Writers & Authors | COO, Blogger, SDR, Member, Mr, Translator, Virtual Assistant, Head of Growth, Author, Editor |
| Cluster 2: Marketing & Sales | Marketing Manager, Digital Marketing Manager, Product Manager, Digital Marketing Specialist, Marketing Director, Business Development Manager, Marketing Specialist, Marketing Consultant, Sales Development Representative, Customer Success Manager |
| Cluster 3: Business & Operations | Project Manager, Consultant, Manager, Account Executive, Student, Business Owner, Operations Manager, Account Manager, Customer Service Representative, Assistant Manager |
| Cluster 4: Executives & Founders | CEO, Owner, Founder, Director, Co-Founder, President, Chief Executive Officer, Managing Director, General Manager, CTO |
| Cluster 5: Technical Roles | Software Engineer, Software Developer, Senior Software Engineer, Web Developer, Data Scientist, Data Analyst, Business Analyst, Data Engineer, Content Writer, Developer |